# Row Hammer Effect and Floating Body Effect of Monolithic 3D Stackable 1T1C DRAM

Sungwon Cho, Po-Kai Hsu, Kiseok Lee, Janak Sharda, Suman Datta, Shimeng Yu
School of Electrical and Computer Engineering, Georgia Institute of Technology, GA, USA

*Abstract*— Monolithic 3D stackable 1T1C DRAM technology is on the rise, with initial prototypes reported by the industry. This work presents a comprehensive reliability study focusing on the intricate interplay between the row hammer effect and the floating body effect. First, using a TCAD model of a 3D DRAM mini-array, we categorize different cases of adjacent cells and show that the notorious row hammer effect induced by charge migration is significantly mitigated compared to 2D DRAM. However, we found that when incorporating an impact ionization model to account for the floating body characteristics of the silicon access transistor, the capacitive coupling between vertically stacked cells is severely exacerbated. Second, we conduct an in-depth investigation into the floating body effect itself. We systematically examine the dependence of this effect on key device parameters, including body thickness, doping concentration, and gate work function.

*Keywords- monolithic, 1T1C, 3D DRAM, row hammer, capacitive coupling, floating body*

## I. Introduction

As traditional DRAM technology faces fundamental scaling challenges below the 10nm node, monolithic stackable 3D DRAM with horizontal 1T1C structures has been proposed as promising solution [1,2]. These prototypes employ Si/SiGe epitaxial growth to fabricate multi-tier horizontal nanosheet access transistors in a bit-cost-scalable manner [1,2]. However, the reliability challenges of this new 3D architecture are not yet comprehensively understood. Specifically, the confined geometry of nanosheet transistors removes bulk access, making the device inherently subject to the floating body effect (FBE), which is expected to be a key concern in 1T1C 3D DRAM [3,4]. It is known that holes generated under a strong electric field can accumulate in the body, leading to amplified leakage current [3,4]. Prior work [3] and [4] analyzed the impact of gate work function and body thickness on the floating body effect. Inspired by observations of barrier lowering [3] and hole accumulation [4], our work applies the classical body effect as framework to systematically analyze how this phenomenon depends on key design parameters and how it interacts with row hammer effect.

Row hammer effect (RHE) has been a well-known reliability/security issue for 2D DRAM, where charge migration through the shared substrate disturbs adjacent cells [5,6]. While the layered geometrical structure of 3D DRAM eliminates this physical path, capacitive coupling between neighboring cells during row hammer attack can persist. RHE-like crosstalk was demonstrated in 1T1C $4F^2$ DRAM [7] and analyzed in capacitor-less 3D DRAM [8]. However, RHE in horizontal 1T1C 3D DRAM with interplay from FBE, has not yet been explored in the literature. We investigate this effect for three different adjacent cell topologies with comprehensive transient analysis on 3D DRAM mini array TCAD model.

## II. Device Configurations and Modeling Methods

Fabrication of a five-layer 3D DRAM prototype with gate length ($L_g$) of ~100 nm silicon access transistor was demonstrated in [2]. We built a TCAD model of 3×2×5-layer mini array 3D DRAM with cell capacitors (Fig.1), following the vertical BL architecture in Ref. [2]. To investigate RHE on 3D DRAM, three cases of different adjacent cells simulated in this work are illustrated in Fig. 2 (a), (b), and (c), which represent the row hammer along y-y', z-z', and x-x' directions. Prior to RHE/FBE simulation, the cell was charged through transient simulation (Fig. 3(a)) and the following simulation bias conditions are illustrated in Fig. 3(b) and (c). The design parameters of the TCAD model are specified in Fig. 5(a). The $I_d$-$V_g$ characteristics and the ratio of BL capacitance and storage node capacitance are extracted to calculate sense margin in Fig. 5(b), which will later be used to calculate row hammer threshold cycle. The parasitic BL capacitance used in the calculation is from Ref. [9], and assumes 64 layers.

## III. Simulation Results for RHE and FBE

To investigate RHE, the storage node is first charged to "1" through transient simulation, after which WL toggling is applied. Our saddle-fin buried gate 2D DRAM model (Fig.4) shows storage node voltage drop of -0.5mV/cycle (Fig.6(b)). Assuming a 144mV of sense margin required, the result corresponds to a row hammer threshold of 1,120 cycles. In contrast, the 3D DRAM model (Fig.1) with two cells arranged in different neighboring topologies (Fig.2) demonstrate enhanced robustness against aggressor cell interference compared to the 2D DRAM. The voltage drops were 0.05 mV/cycle, 0.01 mV/cycle, 0.006 mV/cycle drop (Fig.7(a) left) for case 1 (Fig.2(a)), case 2 (Fig.2(b)) and case 3 (Fig. 2(c)) respectively. However, when an avalanche model for impact ionization is incorporated in the simulation, case 2 (Fig.2(b)) where two cells are vertically adjacent shows a significantly larger storage node voltage drop (Fig.7(a) right). A comparison between simulations with and without the impact ionization model reveals the cell's vulnerability stemming from its floating body. For further validation, a hypothetically grounded body case was also simulated, which showed suppressed charge loss (Fig.7 (b)), and suppressed body potential rise (Fig.7 (d)) compared to case 2 with floating body under the impact ionization model. The row hammer threshold counts for all cases are summarized in Fig.7(c).

To investigate FBE, BL toggling is applied after storage node is charged to "1". The rise in body potential over time for each design at each time stamp, is illustrated in Fig. 8(a). The greater body potential rise in thicker channels result in larger charge loss (Fig.8(b)). Decreasing doping concentration from $N_A = 10^{18}/cm^3$ (~100 dopants) to $N_A = 10^{16}/cm^3$ (almost intrinsic) (case 3,4) reduces the body effect coefficient and the consequent threshold voltage shift. While increasing gate work function (case 4,5) yields a slight larger threshold voltage shift, its higher initial threshold voltage leads to suppressed charge loss. The impact of varying doping concentration and gate work function are both shown in Fig.8(c). As summarized in Fig.8 (c), decreasing the doping concentration and increasing the gate work function (from case 3 to 5) results in a slight increase in on-current (3.9%). Therefore, reducing the body thickness and doping concentration while simultaneously increasing the gate work function effectively mitigates FBE.

## Motivation: Reliability challenges in the monolithic 3D stackable 1T1C DRAM largely unexplored

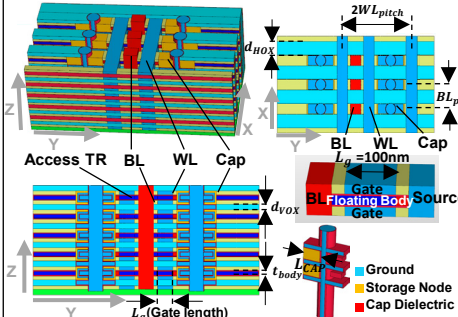

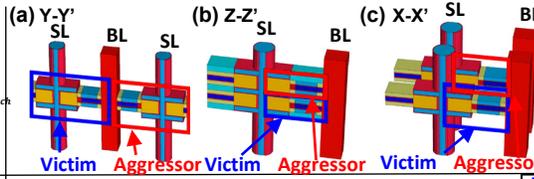

**Fig.2:** 3D DRAM row hammer on different cell-to-cell topologies have not been explored yet. (a) Case1(Y-Y'): Horizontally adjacent cells across bit line (b) Case2(Z-Z'): Vertically adjacent cells across horizontal oxide isolation layer (c) Case3(X-X'): Horizontally adjacent cells across oxide vertical sidewall isolation.

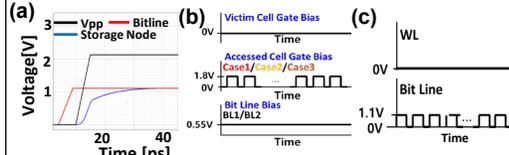

**Fig.1:** 3x2x5 mini array 1T1C Stackable 3D DRAM TCAD model. Bird view, cross-sectional view and top view of the array with geometrically defined access transistors and capacitors.

**Fig.3:** Transient simulation: (a) Write 1 operation on geometrically defined storage node and following simulation bias conditions of (b) Row hammer/ WL coupling effect and (c) Floating body effect.

**Highlights of this work:**
- Demonstration of 3D DRAM TCAD model with a geometrically-defined full cell structure. (An access transistor and a storage node capacitor)
- Investigation of cell-to-cell interference (row hammer effect/ WL coupling) under floating body geometry with a 3x2x5 mini array 3D DRAM TCAD model
- Demonstration of key operations and reliability mechanisms (Write 1, Floating Body Effect, Row Hammer Effect) through comprehensive transient analysis.

## Simulation framework: Characterizing the 3D DRAM model to investigate fundamental reliability challenges

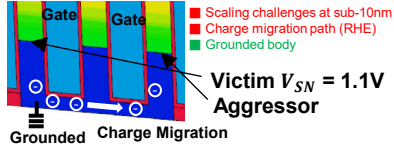

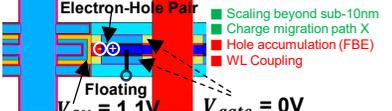

**Fig.4:** Charge migration through shared bulk and grounded body in 2D DRAM vs separated bulk and floating body in 3D DRAM.

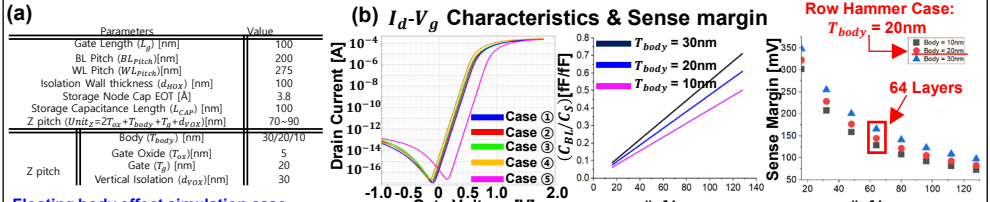

$$\Delta V_{BL} = \frac{1}{(1+C_{BL}/C_S)}\left[\frac{1}{2}V_{DD} - \frac{I_L t_{Ret}}{C_S}\right]$$

Sense margin: 144.14mV @body thickness =20nm, $L_g$=100nm, $N_A = 10^{18}/cm^3$, WF=4.7eV, 64Layers

**Fig.5:** (a) Model parameters and design of simulations implemented in Sentaurus TCAD. Floating body effect explored at different body thickness, doping concentration and gate work function. (b) $I_d - V_g$ characteristics of each case and $C_{BL}/C_S$ extracted to calculate sense margin of Case 2 for future row hammer threshold calculation (BL parasitic analysis is based on Reference [9])

## Row hammer/ WL coupling effect simulation result analysis

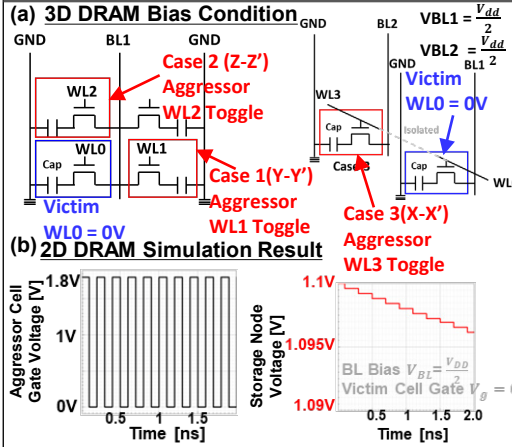

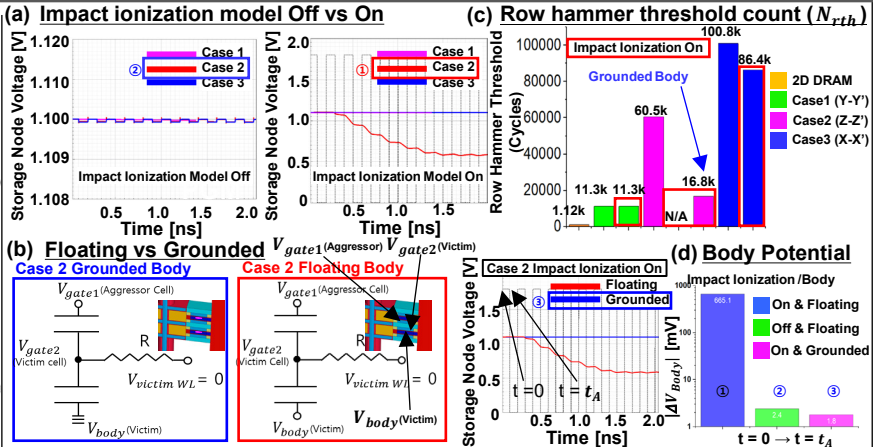

**Fig.6:** (a) Bias condition of row hammer simulation. Victim cell is turned off with BL charged half $V_{DD}$ while 0-$V_{DD}$ pulse is applied on aggressor cell WL. (b) 2D DRAM row hammer simulation result: Aggressor cell gate voltage and storage node charge loss due to charge migration. Result shows row hammer threshold for 2D DRAM within 100mV sense margin is $N_{rth} \leq 1.12k$ cycles.

**Fig.7:** 3D DRAM Row hammer simulation result: (a) Impact ionization model off(Left) vs impact ionization model on(Right): Incorporating the model show significant charge loss for Case 2 (Fig 2-(b)) due to WL coupling effect. (b) Floating and grounded body simulation results show that no bulk access in 3D DRAM may intensify the coupling effect between two vertically adjacent cells. (c) row hammer threshold is compared for each case assuming 144.1mV sense margin from Fig 5-(b) result. (d) Body potential change in Case 2 explains the intensified coupling effect result of "floating and impact ionization model on" (① in Fig 7-(a)) compared to the others (② in Fig 7-(a) and ③ in Fig 7-(b)).

## Floating body effect simulation result analysis

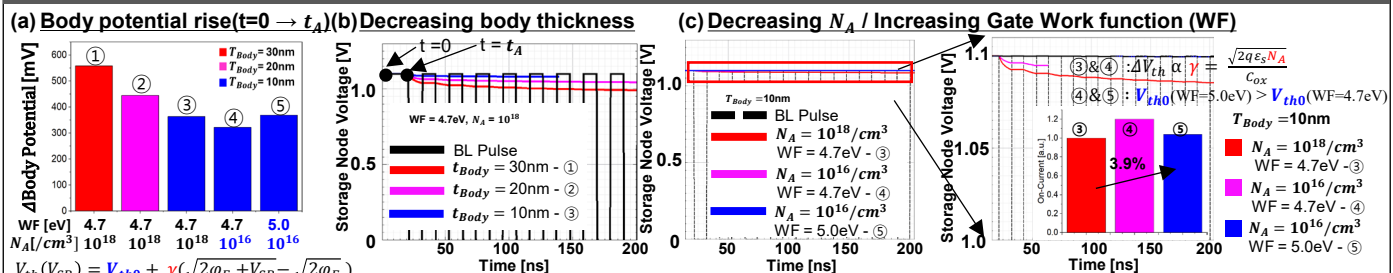

**Fig.8:** Floating body effect: (a) Body potential change for each case. This results to negative threshold voltage shift of the access transistor. (b) Body thickness of 30nm/20nm/10nm at $N_A= 10^{18}$ & Gate work function = 4.7eV (Case ①,②,③ in Fig 5-(a) and Fig 7-(a)) show suppressed charge loss in the smaller body thickness. (c) $N_A = 10^{18}/10^{16}$ at body thickness of 10nm & gate work function = 4.7eV (Case ③,④ in Fig 5-(a) and Fig 7-(a)) show suppressed charge loss in lower doping concentration due to smaller $\Delta V_{th}$. Gate work function = 4.7eV / 5.0eV at body thickness of 10nm & $N_A = 10^{16}$ (Case ④,⑤ in Fig 5-(a) and Fig 7-(a)) show suppressed charge loss in larger gate work function due to larger $V_{th}$ after hole accumulation (Due to larger initial $V_{th}$).